\documentclass[twoside,twocolumn,9pt]{article}
\usepackage{extsizes}
\usepackage[super,sort&compress,comma]{natbib} 
\usepackage[version=3]{mhchem}
\usepackage[left=1.5cm, right=1.5cm, top=1.785cm, bottom=2.0cm]{geometry}
\usepackage{balance}
\usepackage{times,mathptmx}
\usepackage{amssymb}
\usepackage{sectsty}
\usepackage{graphicx} 
\usepackage{lastpage}
\usepackage[format=plain,justification=justified,singlelinecheck=false,font={stretch=1.125,small,sf},labelfont=bf,labelsep=space]{caption}
\usepackage{float}
\usepackage{fancyhdr}
\usepackage{fnpos}
\usepackage[english]{babel}
\addto{\captionsenglish}{%
  
}
\usepackage{array}
\usepackage{droidsans}
\usepackage{charter}
\usepackage[T1]{fontenc}
\usepackage[usenames,dvipsnames]{xcolor}
\usepackage{setspace}
\usepackage[compact]{titlesec}
\usepackage{hyperref}
\usepackage{verbatim}
\usepackage{longtable}

\usepackage{epstopdf}

\newcommand{\CF}{\text{Cl}@\ce{Si32Cl44}}
\newcommand{\CB}{\text{Br}@\ce{Si32Br44}}
\newcommand{\ie}{\textit{i}.\textit{e}.}

\definecolor{cream}{RGB}{222,217,201}

\begin{document}

\pagestyle{fancy}
\thispagestyle{plain}
\fancypagestyle{plain}{

}

\makeFNbottom
\makeatletter
\renewcommand\LARGE{\@setfontsize\LARGE{15pt}{17}}
\renewcommand\Large{\@setfontsize\Large{12pt}{14}}
\renewcommand\large{\@setfontsize\large{10pt}{12}}
\renewcommand\footnotesize{\@setfontsize\footnotesize{7pt}{10}}
\makeatother

\renewcommand{\thefootnote}{\fnsymbol{footnote}}
\renewcommand\footnoterule{\vspace*{1pt}%
\color{cream}\hrule width 3.5in height 0.4pt \color{black}\vspace*{5pt}} 
\setcounter{secnumdepth}{5}

\makeatletter 
\renewcommand\@biblabel[1]{#1}            
\renewcommand\@makefntext[1]%
{\noindent\makebox[0pt][r]{\@thefnmark\,}#1}
\makeatother 
\renewcommand{\figurename}{\small{Fig.}~}
\sectionfont{\sffamily\Large}
\subsectionfont{\normalsize}
\subsubsectionfont{\bf}
\setstretch{1.125} 
\setlength{\skip\footins}{0.8cm}
\setlength{\footnotesep}{0.25cm}
\setlength{\jot}{10pt}
\titlespacing*{\section}{0pt}{4pt}{4pt}
\titlespacing*{\subsection}{0pt}{15pt}{1pt}

\fancyfoot{}
\fancyfoot[LO,RE]{\vspace{-7.1pt}\includegraphics[height=9pt]{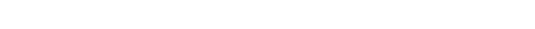}}
\fancyfoot[CO]{\vspace{-7.1pt}\hspace{11.9cm}\includegraphics{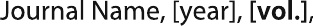}}
\fancyfoot[CE]{\vspace{-7.2pt}\hspace{-13.2cm}\includegraphics{head_foot/RF}}
\fancyfoot[RO]{\footnotesize{\sffamily{1--\pageref{LastPage} ~\textbar  \hspace{2pt}\thepage}}}
\fancyfoot[LE]{\footnotesize{\sffamily{\thepage~\textbar\hspace{4.65cm} 1--\pageref{LastPage}}}}
\fancyhead{}
\renewcommand{\headrulewidth}{0pt} 
\renewcommand{\footrulewidth}{0pt}
\setlength{\arrayrulewidth}{1pt}
\setlength{\columnsep}{6.5mm}
\setlength\bibsep{1pt}

\makeatletter 
\newlength{\figrulesep} 
\setlength{\figrulesep}{0.5\textfloatsep} 

\newcommand{\topfigrule}{\vspace*{-1pt}%
\noindent{\color{cream}\rule[-\figrulesep]{\columnwidth}{1.5pt}} }

\newcommand{\botfigrule}{\vspace*{-2pt}%
\noindent{\color{cream}\rule[\figrulesep]{\columnwidth}{1.5pt}} }

\newcommand{\dblfigrule}{\vspace*{-1pt}%
\noindent{\color{cream}\rule[-\figrulesep]{\textwidth}{1.5pt}} }

\makeatother

\twocolumn[
  \begin{@twocolumnfalse}
  {\includegraphics[height=30pt]{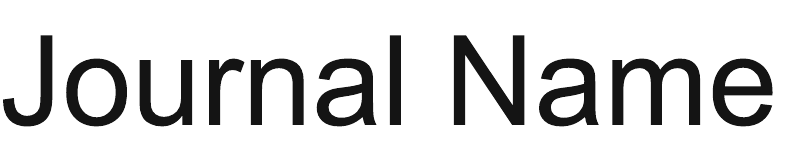}\hfill%
 \raisebox{0pt}[0pt][0pt]{\includegraphics[height=55pt]{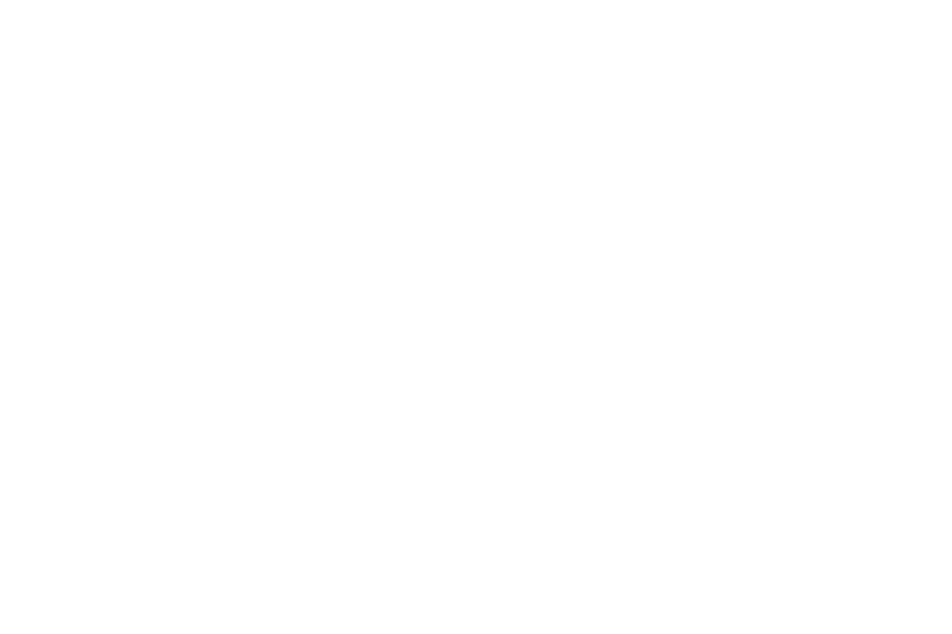}}%
 \\[1ex]%
 \includegraphics[width=18.5cm]{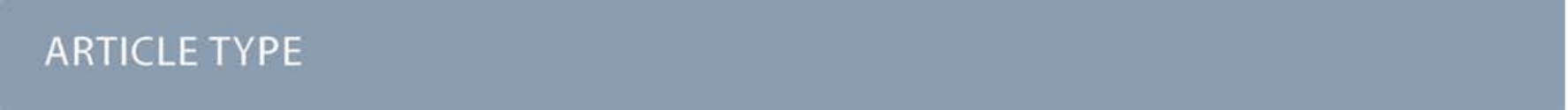}}\par
 \vspace{1em}
\sffamily
\begin{tabular}{m{4.5cm} p{13.5cm} }

\includegraphics{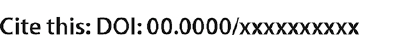} & \noindent\LARGE{\textbf{
Potential energy surface study of X@Si$_{32}$X$_{44}^{-}$(X=Cl, Br) clusters to
decipher the stabilization process of \ce{Si20} fullerene
}} \\
\vspace{0.3cm} & \vspace{0.3cm} \\

 & \noindent\large{Deb Sankar De,$^{\ast}$\textit{$^{a}$} Santanu Saha,\textit{$^{b}$} and Stefan Goedecker\textit{$^{a}$}} \\

\includegraphics{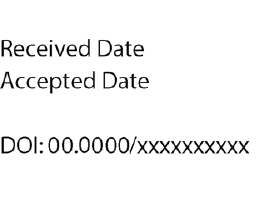} & \noindent\normalsize{
Efforts toward stabilization of the \ce{Si20} fullerene through different schemes have
failed despite several theoretical predictions. However, recently Tillmann {\it et. al.}
succeeded to stabilize the \ce{Si20} fullerene through exohedral decoration with eight
Cl substituents and twelve \ce{SiCl3} groups on the surface and enclosing Cl$^-$ ion.
A deeper understanding on what factors lead to stabilization will open the path for
stabilizing other systems of interest. Here, we employ the minima hopping method within
density functional theory to understand the potential energy surface. The study shows
that the exo-endo halide decoration of the cage alters the glassy nature of the potential
energy surface of pure cage to structure seeker. Further analysis of different properties
of the global minima, reveal that the extra electron instead of residing on the central
encapsulated atom in the cage, it is distributed on the cage and increases the 
encapsulation energy; thereby stabilizing the system. We also provide estimates of the
stability for different kind of exo-endo halide decorations and their feasible realization
in experiments.}


\end{tabular}

\end{@twocolumnfalse} \vspace{0.6cm}

  ]

\renewcommand*\rmdefault{bch}\normalfont\upshape
\rmfamily
\section*{}
\vspace{-1cm}


\footnotetext{\textit{$^{a}$~Department of Physics, Universit\"at Basel, Klingelbergstr. 82, 4056 Basel, Switzerland; E-mail: debsankar.de@unibas.ch; Tel: +41 61 207 3738}}
\footnotetext{\textit{$^{b}$~Institute of Theoretical and Computational Physics, Graz University of Technology, NAWI Graz, 8010 Graz, Austria}}




\section{Introduction}
The discovery of \ce{C60} fullerene in 1985 marks an important milestone in
nano-sciences~\cite{kroto1985c60}. Since then, many experimental and theoretical efforts
had been made to find fullerene structures made out of non-carbon materials 
for various applications~\cite{song2017reversible,yang2015design,johansson2004au32}. Lately, boron based fullerence structure, borospherene
B$_{40}$ was discovered by Zhai et al. in 2014 ~\cite{zhai2014observation}. As both C and Si both are
Group-IV elements without any d orbitals, the Si structures are expected to be similar
to their carbon counterparts. Few examples include linear polysilanes,
silicon nanosheets and nanotubes
~\cite{gibson1985molecular,novoselov2005two,boul1999reversible,miller1989polysilane,okamoto2010silicon}.

Since \ce{C20} forms the smallest known fullerene, \ce{Si20} clusters were widely
studied~\cite{li2000stable}. While C atoms can easily adjust their valence states
to participate single, double and triple bonds, Si strongly favors $sp^{3}$ hybridization 
in connection with single bonds and this inherent constraint leads to the instability
of the \ce{Si20} fullerene~\cite{ho1998structures}. 

Endohedral doping of \ce{Si20} cages with metal atoms, M@\ce{Si20} was believed for a while
to stabilize the cage like geometries. However systematic structure prediction studies show
that this strategy fails~\cite{willand2010structural,sun2002first}. The second approach was
based on the introduction of exhohedral substituents to fully saturate all four silicon
valencies. The \ce{Si20H20} dodecahedron configuration has been identified as the global
minima ~\cite{earley2000ab,zdetsis2007high} in this context. As a third approach,
a combination of both techniques was considered which should lead to a stable M@\ce{Si20H20} 
where M=metal/halide ~\cite{zhang2005structure,pichierri2005encapsulation,palagin2012evaluation}.

The embedding energy of the endohedral dopants in the hydrogenated cages was found to be
smaller than for the bare endohedrally doped fullerene due to weak interaction between the
dopant and the cage~\cite{kumar2007hydrogenated}. Other than that, such surface passivated
endohedrally doped fullerenes proposed in theoretical calculations have very weak interactions
among each other and might therefore be stable building blocks for novel materials.
It was also predicted that the halide ions, especially Br$^{-}$, are ideally suited for the 
synthesis of \ce{Si20H20}~\cite{pichierri2005encapsulation}. However, no such system have been
observed in experiments till date.

In our recent study on \ce{Si20H20}~\cite{de2020nonexistence}, we found that its potential
energy surface(PES) is complex. The spontaneous formation of \ce{Si20H20} by
condensation is unlikely to occur because the time scale for finding the global minimum  is much longer then the time scale for competing processes such as
fragmentation or fusion.

Recently, Tillmann \textit{et. al.}  have developed a one step synthesis protocol by
which the \ce{Si20} dodecahedral core can be stabilized and hence, realized in experiments.
The resulting structure is the $[\CF]^{-}$, which consists of a \ce{Si20} dodecahedral core
with an endohedral Cl$^{-}$ ion and exohedrally decorated with 8 Cl and 12 \ce{SiCl3}
groups~\cite{tillmann2015one}. Their DFT studies on different possible exohedral decoration
with Cl and \ce{SiCl3} of the \ce{Si20} cage structure provides valuable information about
the role of the exohedral decoration on the energetics of the synthesized structure. 
Later, Vargas et. al.\cite{ponce2018stabilizing}  investigated the role of the
central halide ion in stabilizing the system based on energy decomposition and multipole
analysis.

The study of PES can provide estimate of the synthesizability of a desired compound, the barrier
height required to cross and time-scale to synthesize the compound. As explained earlier
in Ref.\cite{de2020nonexistence}, why it is not feasible to synthesize \ce{Si20H20} and opposite
for \ce{C60}. Thus, the PES is equally important as the properties of the synthesized structure. 
The two studies Ref.~\cite{tillmann2015one} and Ref.~\cite{ponce2018stabilizing} have focussed
primarily on the cage structure of \CF and the role of either exohedral or endohedral substituents
in stabilizing the system. However, the role of the PES in making the synthesis of $[\CF]^{-}$
feasible have not been explored yet. Apart from PES, the studies do not address the impact of
different combination of exohedral-endohedral decoration of \ce{Si20} cage and the 
extra electron in its stabilization.

In this work, the minima hopping method~\cite{goedecker2004minima} was used to explore the
potential energy surface(PES) of \ce{Si20} with different endohedral-exohedral decorations.
The runs generated in total 500 distinct structures, which can be classified into fragmented
and non-fragmented structures. Distance-energy plots constructed
for these structures were used to understand the nature of the PES. In addition, 
a detailed analysis of the charge density, stability and reactivity of the ground
state configuration was conducted. Based on a model of the central halide ion trapped in a
spherical potential well, we elucidate the role of exo-endo decorations along with
negative charge in the stabilization process.

\section{Computational Methodology:}
In order to explore the potential energy surface(PES) of the system of interest, the 
minima hopping method (MHM) was employed~\cite{goedecker2004minima,schaefer2015stabilized}.
The PES is explored by performing consecutive short molecular dynamics escape steps followed
by local geometry relaxations, thereby exploiting the Bell-Evans-Polanyi
principle~\cite{roy2008bell}. The relaxed structures are accepted or rejected based on the
threshold set on the energy difference.The transformation pathways were found by the minima
hopping guided path search (MHGPS)~\cite{schaefer2014minima}.

To identify the structural difference between two configurations and removing the duplicate
structures, we have used fingerprint method based on overlap matrix constructed from atom
centered Gaussian type orbitals~\cite{sadeghi2013metrics}. More details on the method can be
found in the appendix. 

All the ab-initio simulation have been performed at the level of density 
functional theory (DFT) as implemented in {\sc BigDFT}~\cite{genovese2008daubechies}. 
This code uses Daubechies wavelets as a basis set. The atoms were described by the
soft Goedecker-type norm-conserving pseudopotentials with non-linear core correction 
~\cite{hartwigsen1998relativistic,goedecker1996separable, krack2005pseudopotentials,
willand2013norm} for the Perdew-Burke-Ernzerhof(PBE) exchange correlation
functional~\cite{perdew1996generalized}. The Libxc library~\cite{marques2012libxc} coupled
with {\sc BigDFT} was used to evaluate the PBE functionals. A grid spacing of 0.4 Bohr was
used. Convergence parameters in {\sc BigDFT} were set such that total energy differences were
converged up to 10$^{-4}$ eV and all configurations were relaxed until the maximal force
component on any atom reached the noise level of the calculation, which was of the order of
1 meV/\AA. 

The difference in charge density($\Delta\rho$) and electron localization function(ELF)
~\cite{silvi1994classification} was calculated using the {\sc vasp} code~\cite{kresse1996efficient}.
The atoms were described by the Projector Augmented Wave potentials as provided in {\sc vasp} for
the PBE functional. In order to mimic the systems under study in free boundary condition, a uniform box
of dimension $20\times20\times20$\,~\AA$^3$ was used. This is the minimum box size required to prevent
self interaction between the clusters. For geometric optimization, a energy cut-off of 350 eV was
used and for ELF calculation 600 eV.


\section{Results and Discussion:}
\begin{figure}[htb!]
\includegraphics[width=1.0\columnwidth,angle=0]{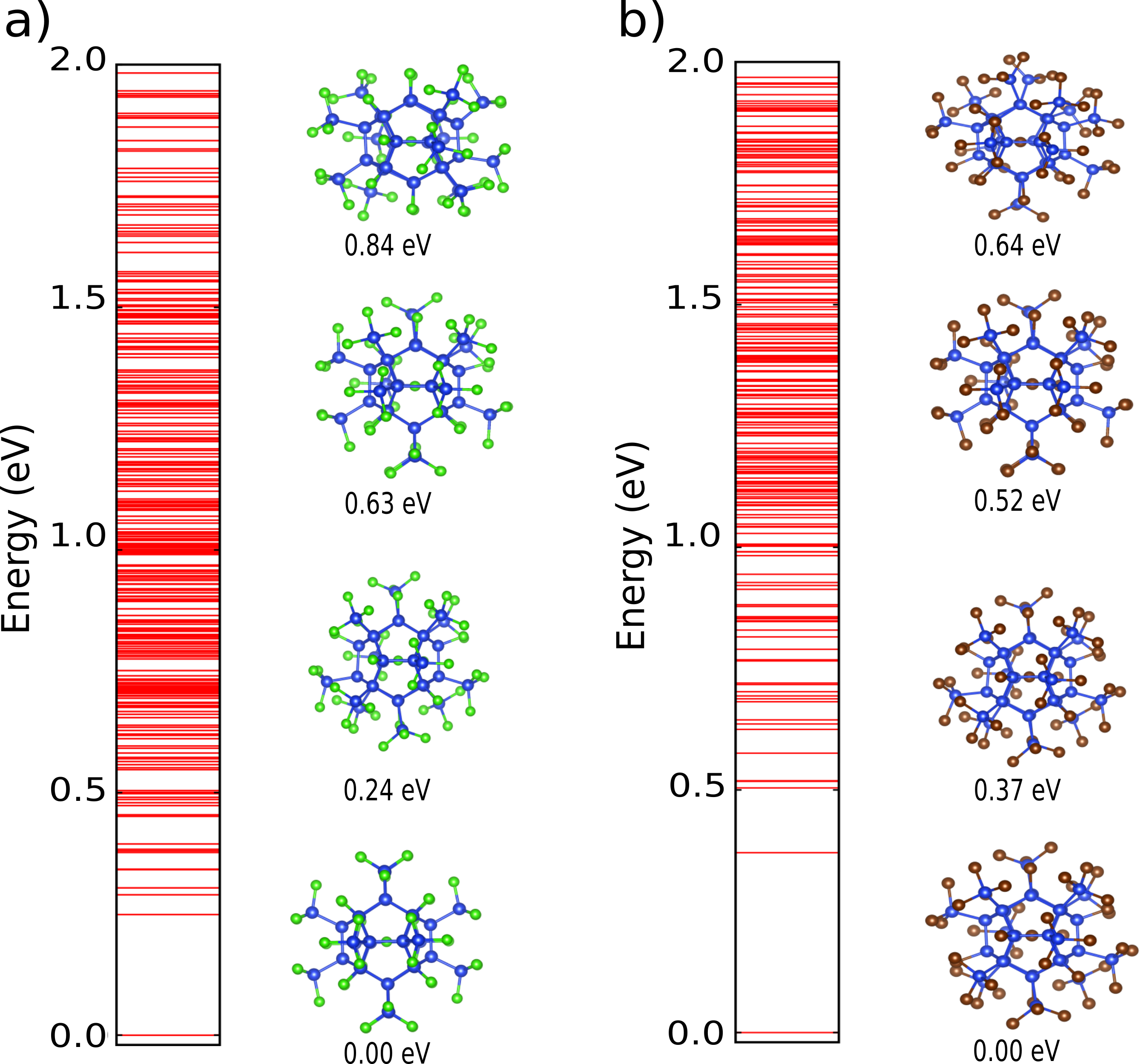}
\caption{The configurational energy spectrum of (a) [\CF]$^{-}$ and 
(b) [\CB]$^{-}$, along with the figures of the ground state configuration
and next 3 lowest energy structures. The Si, Cl and Br atoms are shown by
blue, green and brown sphere, respectively.} 
\label{fig:enespectrum}
\end{figure}

\subsection{Structure, Geometry and Energetics}
A natural starting point of our investigation is the [\CF]$^{-}$ symmetric cage structure, 
experimentally realized by Tillmann et. al.~\cite{tillmann2015one} in 2015. This [\CF]$^{-}$ 
symmetric structure with D$_2$h symmetry is made of a \ce{Si20} dodecaheadral core consisting 
of 20 vertices forming 12 pentagons. Among the 20 vertices, 12 are capped by SiCl$_3$ groups 
arranged in pairs and forming the vertices of an octahedron and the remaining 8 are
capped with Cl atom. The structure also has an endohedrally doped Cl$^{-}$ ion. The Si atoms
of the cage capped with Cl are represented as \ce{Si0} and \ce{SiCl3} decorated Si as \ce{Si1}.

To rapidly perform an exhaustive configurational search of the PES of [\CF]$^{-}$, we carried
out MHM runs starting from different configurations; the reported experimental structure and
few new configurations were created by exchanging the position of the decorative
Cl atoms and \ce{SiCl3} groups. Starting from these different configurations and after removing the
similar structures found in different MHM runs through fingerprint distance, we were left with
$\sim$500 distinct structures. These structures lie in the energy range 0 eV $\leqslant$
$\Delta$E $\leqslant$ 17 eV w.r.t. the lowest energy structure of [\CF]$^{-}$.

We also carried out investigation on the PES of $[\CB]^{-}$.
The 500 structures obtained for $[\CF]^{-}$ were used as templates for generating structures
of $[\CB]^{-}$ by replacing the Cl atoms with Br and post-relaxing them. As Cl and Br are
chemically similar, the $[\CB]^{-}$ structures easily relaxed without significant distortion
and are in the energy range 0 eV $\leqslant$ $\Delta$E $\leqslant$ 12 eV w.r.t. the lowest
energy structure of $[\CB]^{-}$.

The low energy structures generally consist of \ce{Si20} dodecahedron with diverse arrangement
of the Cl and \ce{SiCl3} groups on its vertices. However, the high energy structures consist
of mostly fragmented structures along with distorted cage structures. In most cases among the
fragmented structures, the \ce{SiCl3} gets fragmented to form a free SiCl$_2$ with the remaining
Cl attached to the Si atom of the cage. This feature of structural motif vs energetic ordering appear
to be common in both $[\CF]^{-}$ and $[\CB]^{-}$.

Based on this two distinct structural motif, the whole structure set can be classified into
fragmented and non-fragmented structures. The fragmented structures start to appear at $\Delta$E
= 1.48 eV and $\Delta$E = 0.96 eV for $[\CF]^{-}$ and $[\CB]^{-}$ respectively. Above this energy
range, both fragmented and non-fragmented structures can be found. Beyond 
$\Delta$E $\geqslant$ 4 eV, only the fragmented structures can be found in both
[\CF]$^{-}$ and [\CB]$^{-}$.

The low-energy spectrum of $[\CF]^{-}$ and $[\CB]^{-}$, which consists of primarily
non-fragmented structures is shown in Fig.~\ref{fig:enespectrum}(a) and
Fig.~\ref{fig:enespectrum}(b) respectively w.r.t. their corresponding lowest energy structure.
Along with this, their 4 lowest energy structures are shown in Fig.~\ref{fig:enespectrum}.
Among the predicted [\CF]$^{-1}$ clusters, the experimentally observed cage structure with
D$_{2h}$ symmetry was found to be the lowest energy structure, where the central Cl$^{-}$
ion is located at the center of the cage. This theoretical observation also holds true for
the [\CB]$^{-}$ clusters. The distance between the central Cl$^{-}$ ion and Si from the
\ce{Si20} dodecahedron cage is $\sim$3.28-3.39 \AA~ which is comparable to the experimental
values of 3.32-3.38 \AA. In case of the central Br$^{-}$ ion in the [\CB]$^{-}$ cage structure,
the distance is 3.31-3.40~\AA. Thus, the central halide ion to cage distance is similar in both cases.

The other low energy structures have different arrangements of Si$_0$ and Si$_1$ on pentagons
of the cage. Tillmann et. al.\cite{tillmann2015one} through DFT calculation on different
decoration of \ce{Si20} cage with increasing number of \ce{SiCl3} units have found that the
ideal arrangement would be that in the pentagon, two consecutive vertices are to be occupied
by \ce{SiCl3} units, their immediate neighbouring vertices by Cl atom and the remaining
vertice by a \ce{SiCl3} group. 
Deviation from this symmetric  arrangement of
Si$_1$ was found to cost energy of the structures. This is in agreement with our observations.

The second lowest energy isomer of $[\CF]^{-}$ and $[\CB]^{-}$ are 0.25 eV
and 0.37 eV higher than their ground state respectively. This isomer has one non-ideal 
pentagon consisting of 4 \ce{SiCl3} groups and 1 Cl atoms. The third lowest has two non-ideal
pentagons and the fourth lowest with three non-ideal pentagons. Thus, shifting away from the
ideal pentagonal arrangement for Cl and \ce{SiCl3} leads to an increase in energy. However,
with more non-ideal pentagons, the situation gets complicated as there are many possible
decorative arrangements on these pentagons and their relative energetic ordering is not
straight forward to understand. At a certain point in high energy spectrum, the cage structure
starts to release \ce{SiCl2} with the vertices replaced by Cl atoms.  The very high energy
structures have completely destroyed cages and free SiCl$_4$ molecules.

The structures of $[\CF]^{-}$ and $[\CB]^{-}$ are found to have intermediate to
large HOMO-LUMO gaps. Our calculations of 10 lowest energy structures of
$[\CF]^{-}$ and $[\CB]^{-}$ at the level of semilocal DFT are found to be in the
range 1.7-2.0 eV and 2.0-2.4 eV respectively. No specific trends have been found
among the HOMO-LUMO gap of different structures. The HOMO-LUMO gaps can be found in 
the SI.

\begin{figure}[htb!]
\includegraphics[width=1.0\columnwidth,angle=0]{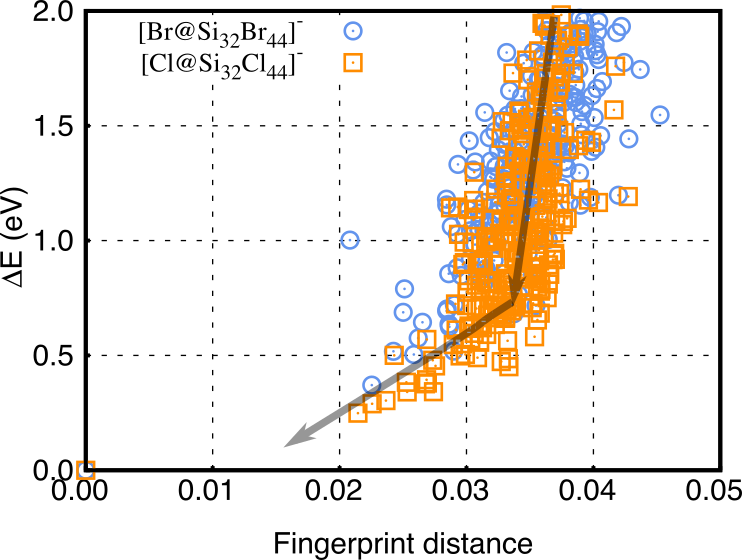}
\caption{
Distance energy (DE) plot of $[\CF]^{-}$(open orange square) and $[\CB]^{-}$(open blue circle).
The energy and the fingerprint distance is w.r.t the lowest energy configuration for each case.
For any structure one can find another structure that is more similar to the
ground state and that is in nearly all cases lower or in a few cases slightly higher in energy. Hence,
on gains quasi continuously energy by moving toward the ground state. The arrow is meant as a guide
to the eye, indicating the average driving force.}
\label{fig:DE-plot}
\end{figure}

\subsection{Fingerprint Distance-Energy}
Apart from the energy ranking of the predicted structures, we also tried to seek insight
into the nature of the PES through fingerprint-distance(DE) vs energy~\cite{de2014glassy} plots.
In such a plot, the energy difference($\Delta$E) is plotted against the fingerprint distance
w.r.t. the ground state structure.

The DE plots of both $[\CF]^{-}$(orange open square) and $[\CB]^{-}$(blue open
circles) as shown in Fig.\ref{fig:DE-plot} are found to be similar as the energy
separation increases steeply with the increase in fingerprint distance. Their
structure seeker character can be easily deduced from the fact that the first few
metastable structures are much higher in energy than the ground state, but not too far
in the configurational distance. This suggests that the barrier for jumping from the
first metastable structure into the ground state is relatively small and that there is,
in general, a strong driving force towards the ground state. The fragmented structures
have been removed while constructing the distance-energy plot.

In contrast, the structures of \ce{Si20} are found to have small $\Delta$E
w.r.t. the putative global minima as shown in SI.
Details about the generation of \ce{Si20} structures are provided in the SI.
Also, \ce{Si20} structures with similar energies are found with both small and large
fingerprint distance. This points to the fact that one can easily get trapped in one
of the local minima and would need to cross barriers of different heights to fall into
the global minima. These features are characteristics of a glassy system. This comparative analysis
reveals the interesting fact that the exo-endo decoration of \ce{Si20} not only stabilizes
the dodecahedron cage but also alters the PES in such a manner that it changes from
glassy to structure seeker.

\begin{figure}[htb!]
\includegraphics[width=1.0\columnwidth,angle=0]{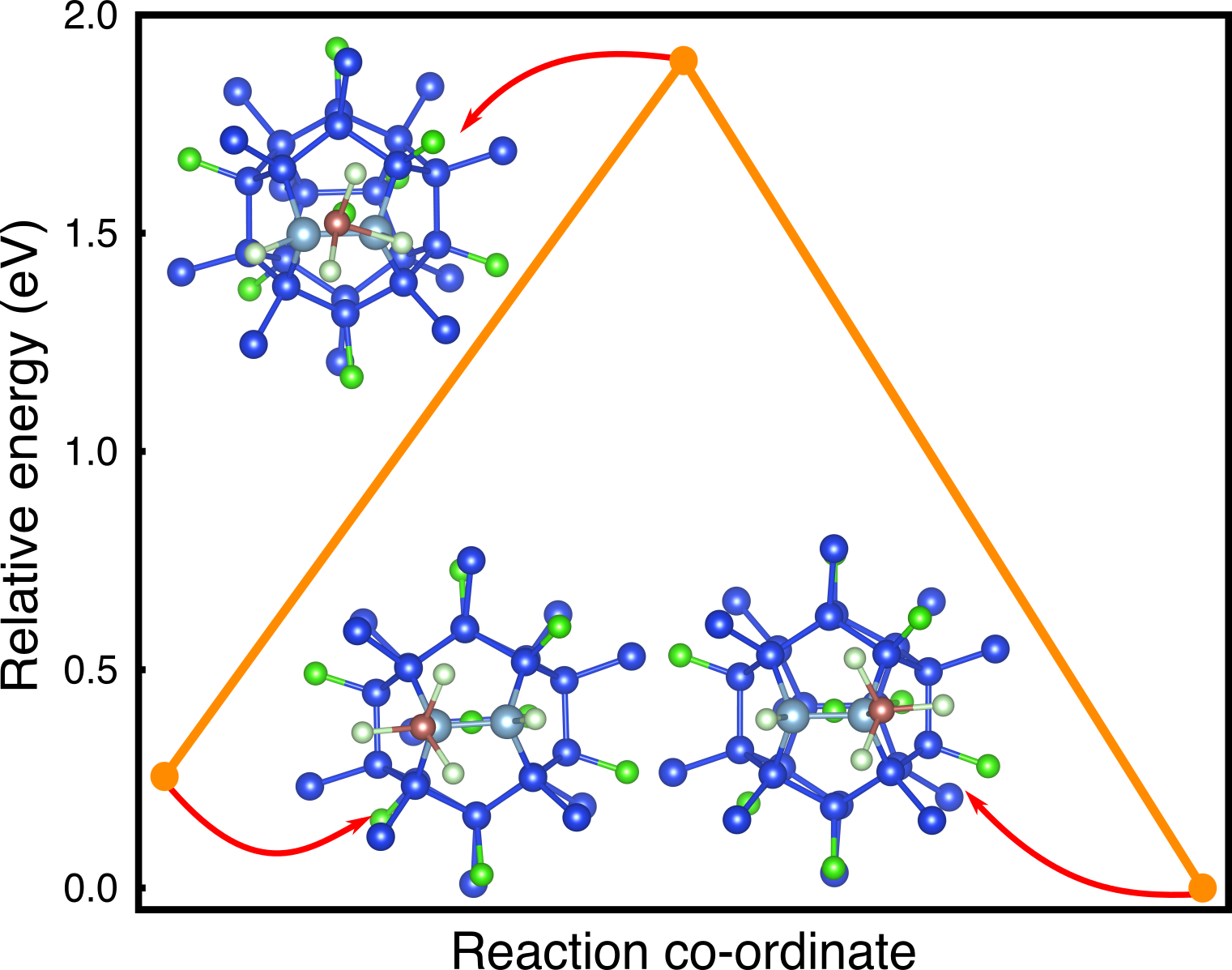}
\caption{The transformation path from the first metastable structure of $[\CF]^{-}$
to the $[\CF]^{-}$ ground state.
Dark blue sphere represent the Si atoms and light blue, the Si atoms of \ce{Si20} cage
undergoing change in their decoration. The Cl atoms of \ce{SiCl3} are omitted
for better visualization. The Si atom of \ce{SiCl3} swapping to another Si atom of
\ce{Si20} cage during the transformation path is shown by pink sphere. Rest all the Cls
are by green atoms.
} 
\label{fig:transition}
\end{figure}

\subsection{Transition State}
The downhill barrier for transiting from the first metastable state to ground state of 
$[\CF]^{-}$ is about 1.6 eV as shown in Fig.\ref{fig:transition}. The barrier height
is quite high compared to k$_B$T at room temperature (which corresponds to 0.027 eV).
The uphill and downhill barriers are similar with a difference of 0.4 eV. In this 
process, the \ce{SiCl3} dissociate into \ce{SiCl2} and Cl. The Cl atom stays on top of Si 
and the \ce{SiCl2} migrates to the nearest Si$_0$ site and forms \ce{SiCl3}.
Standard transition state theory~\cite{eyring1935activated} gives an attempt frequency of
$\omega_A = k_B T/\hslash =  6 \times 10^{12} ~sec^{-1}$
at room temperature. Hence the time required to cross the uphill barrier is 10$^{15}$ sec
at room temperature.

In order to asses the possibility and reaction time at higher temperature, we estimated 
the Boltzmann probabilities as a function of temperature for the 10 lowest energy structures
of $[\CF]^{-}$ and $[\CB]^{-}$. The details of the calculation of the Boltzmann probability 
plot and the figure (Fig. 1 in appendix) is provided in appendix. The Boltzmann probabilities
show that the ground state structure of $[\CB]^{-}$ is dominant until
$\sim$1200 K and for $[\CF]^{-}$ until $\sim$800 K. At higher temperatures,
other low energy structures start to emerge.

Hence, considering 800 K as the upper limit, the time required for transition at 800 K is
around 0.2 msec. Since, the experimental synthesis was carried out in a solvent medium
the actual barrier height could be considerably lower than our calculated height.

\subsection{Bonding and ELF Analysis}

\begin{figure}[htb!]
\includegraphics[width=1.0\columnwidth,angle=0]{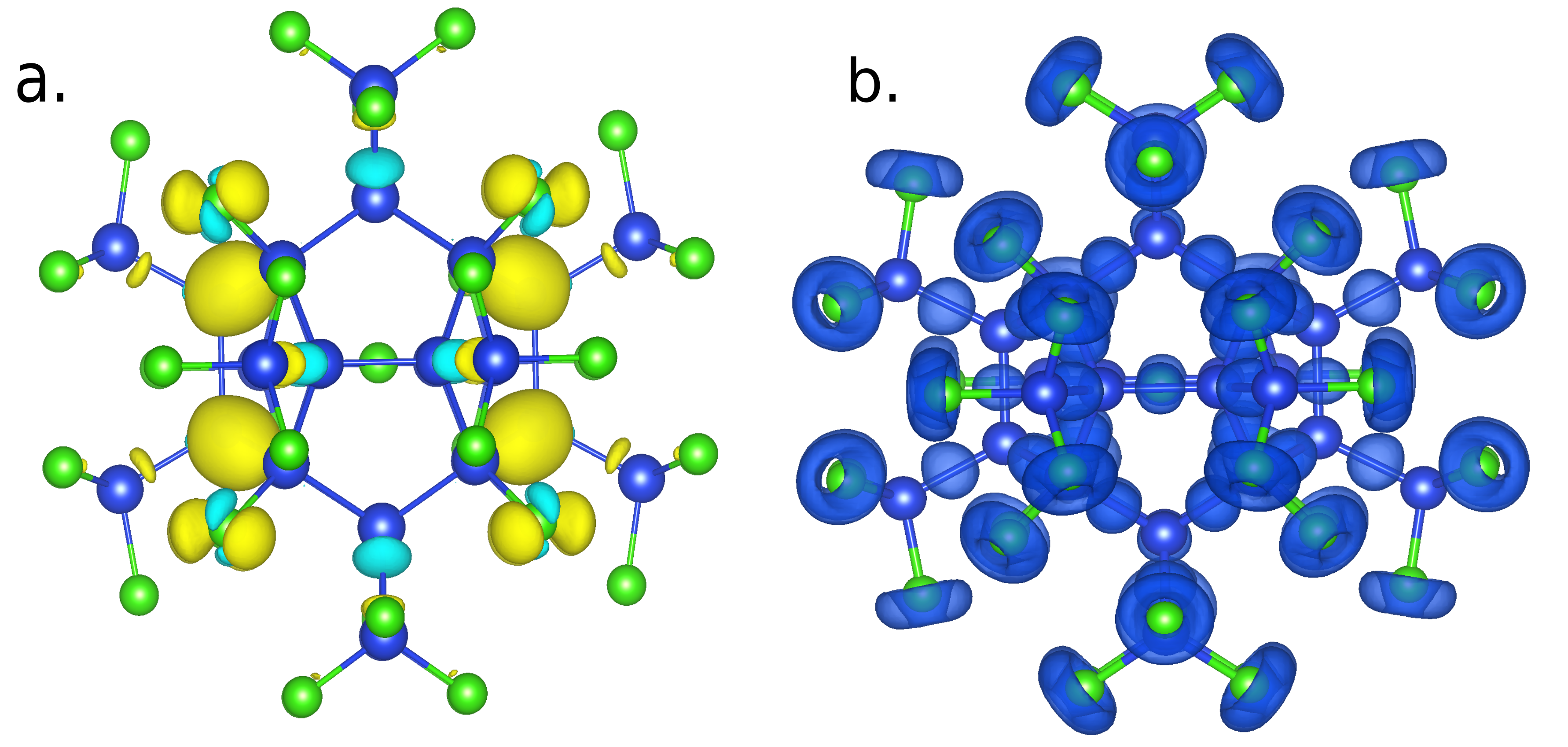}
\caption{a) Charge density difference between neutral and charged \CF. b) ELF of [\CF]$^{-}$ at $\eta$=0.88 . The Si and Cl atoms are shown by blue and green sphere, respectively.} 
\label{fig:probe-atom-ELF}
\end{figure}




In order to investigate where the extra electron added to the [\CF] is localized,
the $\Delta\rho$ between the $[\CF]^{-}$  and [\CF] is shown in
Fig~\ref{fig:probe-atom-ELF}(a) where, the yellow colour indicates the presence of
extra electronic charge and blue electron deficiency. The plot shows that when an electron
is added, it doesn't reside on the central Cl atom. Rather, the Si as well as Cl atoms of the
\ce{Si0} decoration on the cage gain some electronic charge at the expense of losing
electronic charge from Si atoms with \ce{Si1} decoration. There is no significant difference
in the charge density of the Cl in the \ce{SiCl3} groups in \ce{Si1} decoration.

The ELF of $[\CF]^{-}$ at isosurface value of 0.88 is shown in Fig~\ref{fig:probe-atom-ELF}(b).
The circular lobes on the Cl atoms indicate the presence of lone pair electrons on them.

\subsection{Dissociation and Coalescence}


The analysis carried out above on the PES and on ground state configuration of $[\CF]^{-}$
doesn't necessarily shed light on its reactivity. In order to understand its reactive/inert
nature, we accessed if it is stable against dissociation or coalescence.

Fragmentation of $[\CF]^{-}$ has been observed in the laser-desorption ionization(LDI)
MS(-) experiments~\cite{tillmann2015one}. As mentioned earlier, the \ce{SiCl3} group
fragments to form a free \ce{SiCl2} molecule and the remaining Cl atom gets adsorbed on the Si
atom of the \ce{Si20} cage. A major fragmentation cascade is observed due to the extrusions of
such \ce{SiCl2} groups.  Similar structures also were found in our MHM search.
Through our DFT calculation, we found that the ground state configuration is found to be
local minimum and stable under dissociation of one \ce{SiCl2}  and two \ce{SiCl2} 
groups with dissociation energy 1.49 and 2.70 eV, respectively. 
The dissociation energy is defined as the difference between the cluster and the individual 
components of the fragmented parts of the cluster under vacuum condition.

The issue of coalescence of the ground minimum of $[\CF]^{-}$ and $[\CB]^{-}$ was
addressed by estimating the energy required to form dimers. The dimers were created
by bringing the ground state configuration in proximity for three different relative
orientation (see appendix) and 
relaxing them without any constraint. 
It turns out that both the $[\CF]^{-}$ and $[\CB]^{-}$ are chemically inert and 
repel each other, as indicated by their positive binding energy of dimer.
The binding energy of the dimer is defined as the difference between the energy of
the dimer and the energy of the monomers. The factor leading to repulsion is
the presence of lone pair electrons on peripheral Cl atoms as shown in ELF analysis(Fig.~\ref{fig:probe-atom-ELF}).


\subsection{Exohedral and Endohedral Substituents}
The complex dependency of different exohedral and endohedral decorations along with the extra
electron in stabilizing the \ce{Si20} cage makes their understanding challenging. 
Vargas et. al. ~\cite{ponce2018stabilizing} through energy decomposition analysis
tried to decipher the role of substituents and suggested alternative halide based substituents
which can stabilize the cage. Though such analysis are useful, they are complicated and often not
easy to decipher.

The central halide ion in the cage can be considered to be a particle trapped in a spherical
potential well. 
Considering this simple picture, one can easily assimilate the combined effect of the
exohedral/endohedral substituents and the additional electron to a single quantity i.e.,
the curvature of the potential.

For our tests we considered [Cl@Si$_{32}$Cl$_{44}$]$^{-1}$, [Cl@Si$_{32}$Cl$_{44}$],
[Br@Si$_{32}$Cl$_{44}$]$^{-1}$, [Cl@Si$_{20}$H$_{20}$]$^{-1}$ and [Br@Si$_{32}$Br$_{44}$]$^{-1}$
cases. 
The radius
of the cage is different for different decorations. 
The change in energy $\Delta$E(eV) w.r.t. displacement(\AA) of the central atom of the cage
shows that the potential of the charged and neutral case are on top of each other for
all the systems, giving clear indication that the extra electron on the structure
does not effect the stability of the center atom (see appendix). 
The curvature of the systems with 
central Br ion is larger than the ones with central Cl ion. This is due to the large 
ionic radius of Br as compared to Cl confined in the cage structure of similar diameter.
The large ionic radius leads to steep increase in potential w.r.t. displacement. Hence,
the potential is influenced by the substituents and not by the extra electron.

As evident from above, the spring constant K of the potential for [\CF] remains
approximately same for both the neutral and charged system. However, the encapsulation
energy(E.E.) increases by factor of $\sim$1.84 for charged system as compared
to neutral case. The extra stabilization due to addition of an electron
is also supported by the large electron affinity. This behavior holds true for other
endo-exo decoration of \ce{Si20} dodecahedron considered. The electron affinity
depends only on the exohedral substituents. The values for different decorations are
provided in the SI. Thus, the extra electron doesn't lead to increase in interaction
between the central atom and the cage, rather goes to the cage and stabilizes it.

\section{Conclusion} 
In conclusion, through exploration of the PES of $[\CF]^{-}$ and $[\CB]^{-}$, we found that
the experimentally observed symmetric cage structure is the putative global minima. The
exo-endo decoration of \ce{Si20} cage with Cl and \ce{SiCl3} transforms the nature of PES
from glassy to structure seeker. At room temperature the symmetric cage structure is the
most viable isomer during synthesis. However, at higher temperature, other isomers would
start to emerge. The ground state minimum is stable against dissociation and coalescence.
Our stability analysis suggests that the experimental efforts could be oriented towards the
synthesis of the hypothetical counterpart $[\CB]^{-}$ of $[\CF]^{-}$.

The $\Delta\rho$ shown in Fig.\ref{fig:probe-atom-ELF} coupled with the central atom 
displacement study modeled into a particle trapped in a spherical potential well clearly
indicate that the extra electron instead of residing on the central halide atom, gets distributed
on the cage and stabilizes it by increasing the inter-atomic interaction within the cage.
rather between the central atom and the cage. As a result of this, the extra electron leads
to significant increase in encapsulation energy, thereby making the combined system more stable.

\section*{Conflicts of interest}
 There are no conflicts to declare.

\section*{Acknowledgements}
D.S.D acknowledge support from the Swiss National Science Foundation. 
Computational resources provided by the Swiss National Supercomputing Center (CSCS) in Lugano under the projects s707 and s963 are gratefully acknowledged.  Calculations were also performed at the sciCORE (http://scicore.unibas.ch/) scientific computing core facility at the University of Basel.
This was supported by the NCCR MARVEL, funded by the Swiss National Science Foundation.

\newpage
\appendix

\section{Finger print method:}
The configurational
fingerprints are given by the eigenvalues of an overlap matrix
~\cite{sadeghi}. \begin{equation}
O_{ij}=\int \phi_{i}^{l}{(\boldsymbol r)} \phi_{i}^{l^{\prime}}{(\boldsymbol r)} d{\boldsymbol r}
\end{equation}

where $\phi{i}$ are the Gaussian type orbitals centered on the atom at position $\boldsymbol r_{i}$ and is given by
\begin{equation}
\phi_{i}^{l}{(\boldsymbol r)}\propto (x-x_{i})^{l_{x}} (y-y_{i})^{l_{y}} (z-z_{i})^{l_{z}} \exp(-\alpha_{i}|\boldsymbol r -\boldsymbol r_{i}|)^{2})
\end{equation}
Here $\boldsymbol l$=($l_{x},l_{y}, l_{z}$) is aa angular momentum L=$l_{x}+l_{y}+l_{z}$. The orbitals are classified depending on their value of L \ie $s$-type orbital is L=0, $p$-type orbital is L=1, or $d$-type orbital is L=3. $\alpha_{i}$ is the orbital width and they are inversly proportional to the covalent radius of the atoms on which the orbitals are centered on. The structural difference is given by the root mean square difference of the two fingerprint vectors. This fingerprint distance is invariant under translations, rotations, reflections as well as under the permutation of the atomic indices.

\newpage

\begin{figure}[htb!]
\includegraphics[width=1.0\columnwidth,angle=0]{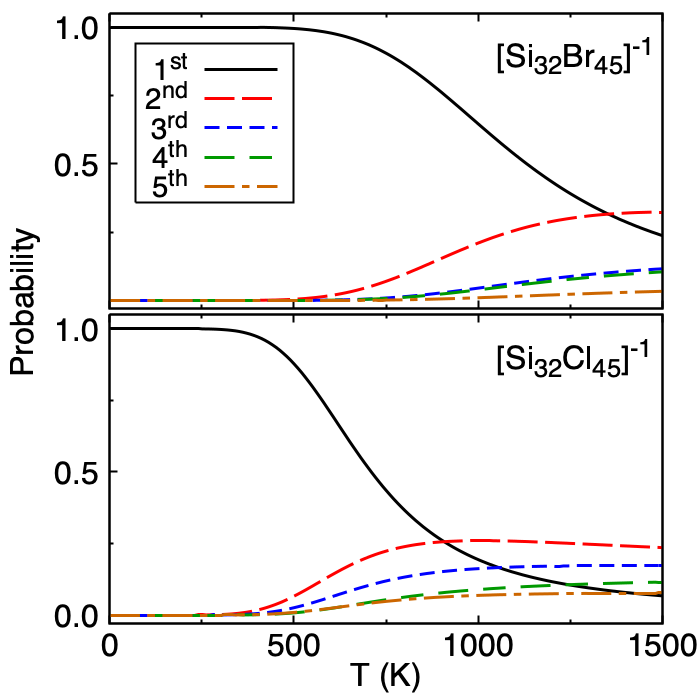}
\caption{The Boltzmann probability distribution of 5 lowest energy structures of [\CF]$^{-}$ 
(bottom panel)and 
The Boltzmann probability have been estimated for 10 lowest energy structures.} 
\label{fig:Boltzprob}
\end{figure}

\newpage

\begin{figure}[htbp!]
\includegraphics[width=1.0\columnwidth,angle=0]{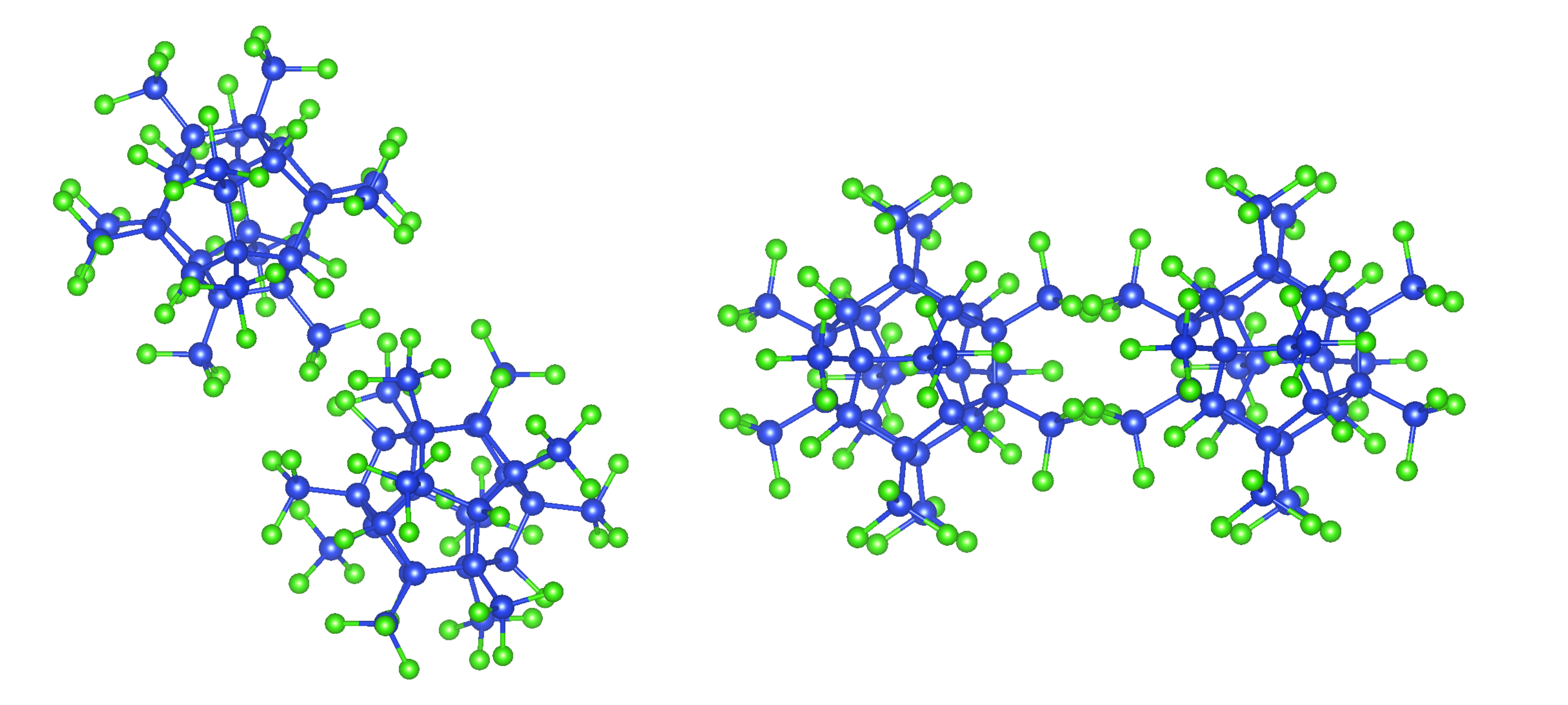}
\caption{Dimers are constructed by adding two GS configurations along X, Z and X-Y directions. Here we show the structures which are formed by adding the GS configurations along X and X-Y direction.  } 
\label{fig:minima_fig}
\end{figure}

\newpage

\begin{figure}[htb!]
\includegraphics[width=1.0\columnwidth,angle=0]{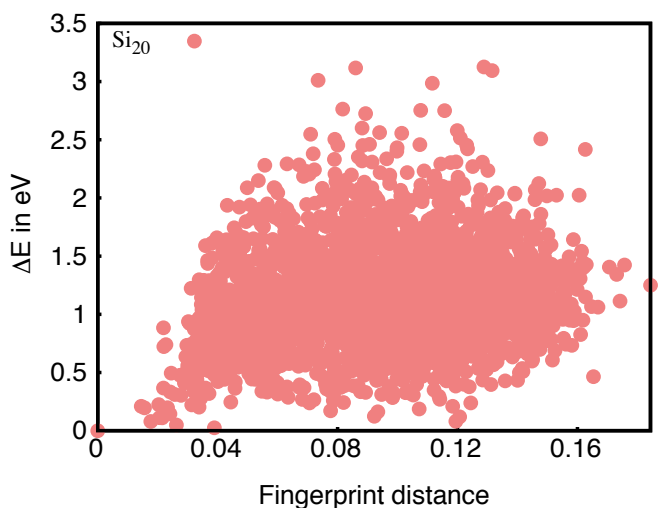}
\caption{Distance Energy (DE) plot of \ce{Si20} clusters. The finger print distances and energy differences are calculated with respect to the ground state configuration. There is no driving force towards the ground state.} 
\label{fig:DE-plot}
\end{figure}

\begin{table}
\begin{tabular}{c c }
  System & Gap in Angstroem \\ \hline \hline
\ce{Si32Cl45}$^{-}$ & 6.736 \\
\ce{Si32Cl45} & 6.787   \\
\ce{Si32Cl44Br}$^{-}$  & 6.749  \\ 
\ce{Si32Br45}$^{-}$  & 6.679  \\ 
\ce{Si20H20Cl}$^{-}$  & 6.629  \\ 

\end{tabular}  
 \caption{Diameter of the gap in the X-Y plane.}
  \label{tab:si32cl45Disen}
\end{table}

\section{Exohedral and Endohedral Substituents}
Based on this simplistic picture of a
particle trapped in a 1D-potential, we estimated the spring constant of the quadratic potential. The potential
can be easily constructed by obtaining the energy difference as a function of displacement, where the reference
energy is the energy of the geometrically relaxed structure. During the displacement of the central atom, the 
whole structure is kept fixed and only scf runs were done to obtain the total energy.
In addition we also calculated encapsulation energy (E.E.) and electron affinity (E.A.).
Encapsulation energy is given by: 
\begin{equation}
\text{E.E.} = \text{E$_{system}$}-\text{E$_{cage}$}-\text{E$_{central\ atom}$}
\end{equation}

where E$_{system}$ is the energy of the whole system cage + center anion(neutral or charged),
          E$_{cage}$ energy of the empty cage without the central atom and
          E$_{central\ anion}$ energy of the central halogen (neutral or charged state).
Electron affinity (E.A.) is given by
 \begin{equation}
\text{E.A.} =  \text{E$_{neutral\ system}$} - \text{E$_{charged\ system}$}
\end{equation}
where E$_{charged\ system}$ is the energy of the charged cage and E$_{neutral\ system}$ is the energy of the charged cage.

\begin{figure*}[thb!]
\includegraphics[width=1.0\columnwidth,angle=0]{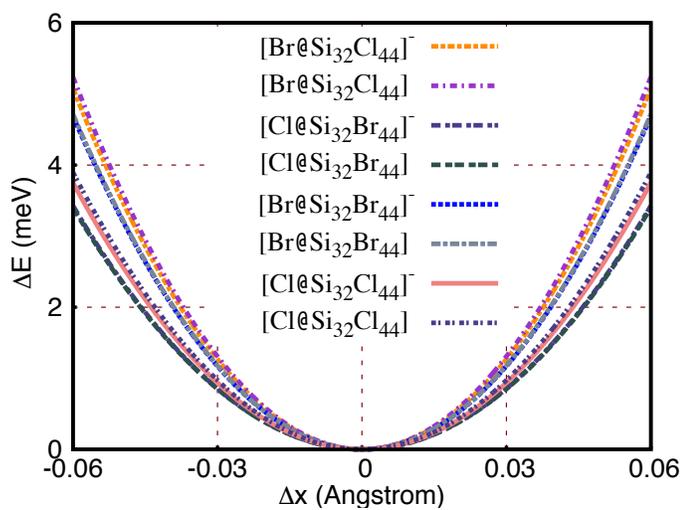}
\caption{Change in total energy $\Delta$E(eV) as a function of displacement ($\Delta$x)
in \AA ~of the center atom for different combination of exohedral and endohedral decoration
in the \ce{Si20} dodecahedron cage. The reference energy for each case was the total energy
of the ground state structure and the reference length for the displacement was the radius
of the cage for corresponding decoration. The cage structure was kept fixed during the 
displacement of the center atom. } 
\label{fig:displacement}
\end{figure*}

The E.E., spring constant K and E.A. of these systems are listed in Table.\ref{tab:BEKstrength}. 
As seen above in the Fig.\ref{fig:displacement} and also from the K values, they are similar for charged and
neutral states. The E.E. sheds light on the role of the extra electron, i.e. it reduces the total energy,
making it more stable and is also supported by the large E.A. Its also interesting to note that the E.A.
is not effected by the type of central halide ion, but rather by the exohedral decoration. 

\newpage

\begin{table}[!htb]
\caption{Encapsulation Energy (eV), Harmonic oscillator strength K (eV/\AA$^2$) and
electron affinity E.A.(eV) for different decorations of Si$_{20}$ dodecahedron.}
\centering
\begin{tabular}{|c|c|c|c|c|c|}
\hline 
System   &  \multicolumn{2}{c|}{E.E. (eV)} &  \multicolumn{2}{c|}{K (eV/\AA$^2$)} & E.A.  \\
\cline{2-5} 
         &  Neutral & Charge & Neutral & Charge & (eV)  \\
\hline
Cl@Si$_{32}$Cl$_{44}$ & -2.34  &  -4.38 & 2.16  & 2.08(X,Y,Z) & 5.68  \\
\hline
Br@Si$_{32}$Cl$_{44}$ &  -1.87 & -4.15  & 2.91 & 2.82 & 5.68  \\
\hline
Cl@Si$_{32}$Br$_{44}$ &  -2.48 & -4.28 & 1.89 & 1.90  & 5.43  \\
\hline
Br@Si$_{32}$Br$_{44}$ & -2.03  & -4.07  & 2.62 & 2.60 & 5.44  \\
\hline
Cl@Si$_{20}$H$_{20}$ & -0.07  & -0.10  & 2.70 & 2.69 & 0.16  \\
\hline
Br@Si$_{20}$H$_{20}$ & -0.05  & -0.09  & 3.82 & 3.81 & 0.16  \\
\hline
 \end{tabular}
\label{tab:BEKstrength}
\end{table}

\newpage

\balance


\bibliography{Si_manu} 

\providecommand*{\mcitethebibliography}{\thebibliography}
\csname @ifundefined\endcsname{endmcitethebibliography}
{\let\endmcitethebibliography\endthebibliography}{}
\begin{mcitethebibliography}{39}
\providecommand*{\natexlab}[1]{#1}
\providecommand*{\mciteSetBstSublistMode}[1]{}
\providecommand*{\mciteSetBstMaxWidthForm}[2]{}
\providecommand*{\mciteBstWouldAddEndPuncttrue}
  {\def\EndOfBibitem{\unskip.}}
\providecommand*{\mciteBstWouldAddEndPunctfalse}
  {\let\EndOfBibitem\relax}
\providecommand*{\mciteSetBstMidEndSepPunct}[3]{}
\providecommand*{\mciteSetBstSublistLabelBeginEnd}[3]{}
\providecommand*{\EndOfBibitem}{}
\mciteSetBstSublistMode{f}
\mciteSetBstMaxWidthForm{subitem}
{(\emph{\alph{mcitesubitemcount}})}
\mciteSetBstSublistLabelBeginEnd{\mcitemaxwidthsubitemform\space}
{\relax}{\relax}

\bibitem[Kroto \emph{et~al.}(1985)Kroto, Heath, O'Brien, Curl, and
  Smalley]{kroto1985c60}
H.~W. Kroto, J.~R. Heath, S.~C. O'Brien, R.~F. Curl and R.~E. Smalley,
  \emph{nature}, 1985, \textbf{318}, 162--163\relax
\mciteBstWouldAddEndPuncttrue
\mciteSetBstMidEndSepPunct{\mcitedefaultmidpunct}
{\mcitedefaultendpunct}{\mcitedefaultseppunct}\relax
\EndOfBibitem
\bibitem[Song \emph{et~al.}(2017)Song, Querebillo, G{\"o}tz, Katz, Kuhlmann,
  Gernert, Weidinger, and Hildebrandt]{song2017reversible}
W.~Song, C.~Querebillo, R.~G{\"o}tz, S.~Katz, U.~Kuhlmann, U.~Gernert,
  I.~Weidinger and P.~Hildebrandt, \emph{Nanoscale}, 2017, \textbf{9},
  8380--8387\relax
\mciteBstWouldAddEndPuncttrue
\mciteSetBstMidEndSepPunct{\mcitedefaultmidpunct}
{\mcitedefaultendpunct}{\mcitedefaultseppunct}\relax
\EndOfBibitem
\bibitem[Yang \emph{et~al.}(2015)Yang, Ji, Lan, Xu, Liu, and
  Xu]{yang2015design}
Z.~Yang, Y.-L. Ji, G.~Lan, L.-C. Xu, X.~Liu and B.~Xu, \emph{Solid State
  Communications}, 2015, \textbf{217}, 38--42\relax
\mciteBstWouldAddEndPuncttrue
\mciteSetBstMidEndSepPunct{\mcitedefaultmidpunct}
{\mcitedefaultendpunct}{\mcitedefaultseppunct}\relax
\EndOfBibitem
\bibitem[Johansson \emph{et~al.}(2004)Johansson, Sundholm, and
  Vaara]{johansson2004au32}
M.~P. Johansson, D.~Sundholm and J.~Vaara, \emph{Angewandte Chemie}, 2004,
  \textbf{116}, 2732--2735\relax
\mciteBstWouldAddEndPuncttrue
\mciteSetBstMidEndSepPunct{\mcitedefaultmidpunct}
{\mcitedefaultendpunct}{\mcitedefaultseppunct}\relax
\EndOfBibitem
\bibitem[Zhai \emph{et~al.}(2014)Zhai, Zhao, Li, Chen, Bai, Hu, Piazza, Tian,
  Lu, Wu,\emph{et~al.}]{zhai2014observation}
H.-J. Zhai, Y.-F. Zhao, W.-L. Li, Q.~Chen, H.~Bai, H.-S. Hu, Z.~A. Piazza,
  W.-J. Tian, H.-G. Lu, Y.-B. Wu \emph{et~al.}, \emph{Nature chemistry}, 2014,
  \textbf{6}, 727--731\relax
\mciteBstWouldAddEndPuncttrue
\mciteSetBstMidEndSepPunct{\mcitedefaultmidpunct}
{\mcitedefaultendpunct}{\mcitedefaultseppunct}\relax
\EndOfBibitem
\bibitem[Gibson \emph{et~al.}(1985)Gibson, Weagley, Mosher, Kaplan, Prest, and
  Epstein]{gibson1985molecular}
H.~W. Gibson, R.~J. Weagley, R.~A. Mosher, S.~Kaplan, W.~M. Prest and A.~J.
  Epstein, \emph{Phys. Rev. B}, 1985, \textbf{31}, 2338--2342\relax
\mciteBstWouldAddEndPuncttrue
\mciteSetBstMidEndSepPunct{\mcitedefaultmidpunct}
{\mcitedefaultendpunct}{\mcitedefaultseppunct}\relax
\EndOfBibitem
\bibitem[Novoselov \emph{et~al.}(2005)Novoselov, Geim, Morozov, Jiang,
  Katsnelson, Grigorieva, Dubonos, and Firsov]{novoselov2005two}
K.~S. Novoselov, A.~K. Geim, S.~V. Morozov, D.~Jiang, M.~I.~. Katsnelson, I.~V.
  Grigorieva, S.~V. Dubonos and A.~A. Firsov, \emph{Nature}, 2005,
  \textbf{438}, 197--200\relax
\mciteBstWouldAddEndPuncttrue
\mciteSetBstMidEndSepPunct{\mcitedefaultmidpunct}
{\mcitedefaultendpunct}{\mcitedefaultseppunct}\relax
\EndOfBibitem
\bibitem[Boul \emph{et~al.}(1999)Boul, Liu, Mickelson, Huffman, Ericson,
  Chiang, Smith, Colbert, Hauge, Margrave, and Smalley]{boul1999reversible}
P.~Boul, J.~Liu, E.~Mickelson, C.~Huffman, L.~Ericson, I.~Chiang, K.~Smith,
  D.~Colbert, R.~Hauge, J.~Margrave and R.~Smalley, \emph{Chemical Physics
  Letters}, 1999, \textbf{310}, 367--372\relax
\mciteBstWouldAddEndPuncttrue
\mciteSetBstMidEndSepPunct{\mcitedefaultmidpunct}
{\mcitedefaultendpunct}{\mcitedefaultseppunct}\relax
\EndOfBibitem
\bibitem[Miller and Michl(1989)]{miller1989polysilane}
R.~D. Miller and J.~Michl, \emph{Chemical Reviews}, 1989, \textbf{89},
  1359--1410\relax
\mciteBstWouldAddEndPuncttrue
\mciteSetBstMidEndSepPunct{\mcitedefaultmidpunct}
{\mcitedefaultendpunct}{\mcitedefaultseppunct}\relax
\EndOfBibitem
\bibitem[Okamoto \emph{et~al.}(2010)Okamoto, Kumai, Sugiyama, Mitsuoka,
  Nakanishi, Ohta, Nozaki, Yamaguchi, Shirai, and Nakano]{okamoto2010silicon}
H.~Okamoto, Y.~Kumai, Y.~Sugiyama, T.~Mitsuoka, K.~Nakanishi, T.~Ohta,
  H.~Nozaki, S.~Yamaguchi, S.~Shirai and H.~Nakano, \emph{Journal of the
  American Chemical Society}, 2010, \textbf{132}, 2710--2718\relax
\mciteBstWouldAddEndPuncttrue
\mciteSetBstMidEndSepPunct{\mcitedefaultmidpunct}
{\mcitedefaultendpunct}{\mcitedefaultseppunct}\relax
\EndOfBibitem
\bibitem[Li and Cao(2000)]{li2000stable}
B.-x. Li and P.-l. Cao, \emph{Physical Review A}, 2000, \textbf{62},
  023201\relax
\mciteBstWouldAddEndPuncttrue
\mciteSetBstMidEndSepPunct{\mcitedefaultmidpunct}
{\mcitedefaultendpunct}{\mcitedefaultseppunct}\relax
\EndOfBibitem
\bibitem[Ho \emph{et~al.}(1998)Ho, Shvartsburg, Pan, Lu, Wang, Wacker, Fye, and
  Jarrold]{ho1998structures}
K.-M. Ho, A.~A. Shvartsburg, B.~Pan, Z.-Y. Lu, C.-Z. Wang, J.~G. Wacker, J.~L.
  Fye and M.~F. Jarrold, \emph{Nature}, 1998, \textbf{392}, 582--585\relax
\mciteBstWouldAddEndPuncttrue
\mciteSetBstMidEndSepPunct{\mcitedefaultmidpunct}
{\mcitedefaultendpunct}{\mcitedefaultseppunct}\relax
\EndOfBibitem
\bibitem[Willand \emph{et~al.}(2010)Willand, Gramzow, Ghasemi, Genovese,
  Deutsch, Reuter, and Goedecker]{willand2010structural}
A.~Willand, M.~Gramzow, S.~A. Ghasemi, L.~Genovese, T.~Deutsch, K.~Reuter and
  S.~Goedecker, \emph{Physical Review B}, 2010, \textbf{81}, 201405\relax
\mciteBstWouldAddEndPuncttrue
\mciteSetBstMidEndSepPunct{\mcitedefaultmidpunct}
{\mcitedefaultendpunct}{\mcitedefaultseppunct}\relax
\EndOfBibitem
\bibitem[Sun \emph{et~al.}(2002)Sun, Wang, Briere, Kumar, Kawazoe, and
  Jena]{sun2002first}
Q.~Sun, Q.~Wang, T.~Briere, V.~Kumar, Y.~Kawazoe and P.~Jena, \emph{Physical
  Review B}, 2002, \textbf{65}, 235417\relax
\mciteBstWouldAddEndPuncttrue
\mciteSetBstMidEndSepPunct{\mcitedefaultmidpunct}
{\mcitedefaultendpunct}{\mcitedefaultseppunct}\relax
\EndOfBibitem
\bibitem[Earley(2000)]{earley2000ab}
C.~W. Earley, \emph{The Journal of Physical Chemistry A}, 2000, \textbf{104},
  6622--6627\relax
\mciteBstWouldAddEndPuncttrue
\mciteSetBstMidEndSepPunct{\mcitedefaultmidpunct}
{\mcitedefaultendpunct}{\mcitedefaultseppunct}\relax
\EndOfBibitem
\bibitem[Zdetsis(2007)]{zdetsis2007high}
A.~D. Zdetsis, \emph{Physical Review B}, 2007, \textbf{76}, 075402\relax
\mciteBstWouldAddEndPuncttrue
\mciteSetBstMidEndSepPunct{\mcitedefaultmidpunct}
{\mcitedefaultendpunct}{\mcitedefaultseppunct}\relax
\EndOfBibitem
\bibitem[Zhang \emph{et~al.}(2005)Zhang, Wu, and Jiao]{zhang2005structure}
C.-Y. Zhang, H.-S. Wu and H.~Jiao, \emph{Chemical physics letters}, 2005,
  \textbf{410}, 457--461\relax
\mciteBstWouldAddEndPuncttrue
\mciteSetBstMidEndSepPunct{\mcitedefaultmidpunct}
{\mcitedefaultendpunct}{\mcitedefaultseppunct}\relax
\EndOfBibitem
\bibitem[Pichierri \emph{et~al.}(2005)Pichierri, Kumar, and
  Kawazoe]{pichierri2005encapsulation}
F.~Pichierri, V.~Kumar and Y.~Kawazoe, \emph{Chemical physics letters}, 2005,
  \textbf{406}, 341--344\relax
\mciteBstWouldAddEndPuncttrue
\mciteSetBstMidEndSepPunct{\mcitedefaultmidpunct}
{\mcitedefaultendpunct}{\mcitedefaultseppunct}\relax
\EndOfBibitem
\bibitem[Palagin and Reuter(2012)]{palagin2012evaluation}
D.~Palagin and K.~Reuter, \emph{Physical Review B}, 2012, \textbf{86},
  045416\relax
\mciteBstWouldAddEndPuncttrue
\mciteSetBstMidEndSepPunct{\mcitedefaultmidpunct}
{\mcitedefaultendpunct}{\mcitedefaultseppunct}\relax
\EndOfBibitem
\bibitem[Kumar and Kawazoe(2007)]{kumar2007hydrogenated}
V.~Kumar and Y.~Kawazoe, \emph{Physical Review B}, 2007, \textbf{75},
  155425\relax
\mciteBstWouldAddEndPuncttrue
\mciteSetBstMidEndSepPunct{\mcitedefaultmidpunct}
{\mcitedefaultendpunct}{\mcitedefaultseppunct}\relax
\EndOfBibitem
\bibitem[De \emph{et~al.}(2020)De, Schaefer, von Issendorff, and
  Goedecker]{de2020nonexistence}
D.~S. De, B.~Schaefer, B.~von Issendorff and S.~Goedecker, \emph{Physical
  Review B}, 2020, \textbf{101}, 214303\relax
\mciteBstWouldAddEndPuncttrue
\mciteSetBstMidEndSepPunct{\mcitedefaultmidpunct}
{\mcitedefaultendpunct}{\mcitedefaultseppunct}\relax
\EndOfBibitem
\bibitem[Tillmann \emph{et~al.}(2015)Tillmann, Wender, Bahr, Bolte, Lerner,
  Holthausen, and Wagner]{tillmann2015one}
J.~Tillmann, J.~H. Wender, U.~Bahr, M.~Bolte, H.-W. Lerner, M.~C. Holthausen
  and M.~Wagner, \emph{Angewandte Chemie}, 2015, \textbf{127}, 5519--5523\relax
\mciteBstWouldAddEndPuncttrue
\mciteSetBstMidEndSepPunct{\mcitedefaultmidpunct}
{\mcitedefaultendpunct}{\mcitedefaultseppunct}\relax
\EndOfBibitem
\bibitem[Ponce-Vargas and Mu\~{n}oz Castro(2018)]{ponce2018stabilizing}
M.~Ponce-Vargas and A.~Mu\~{n}oz Castro, \emph{The Journal of Physical
  Chemistry C}, 2018, \textbf{122}, 12551--12558\relax
\mciteBstWouldAddEndPuncttrue
\mciteSetBstMidEndSepPunct{\mcitedefaultmidpunct}
{\mcitedefaultendpunct}{\mcitedefaultseppunct}\relax
\EndOfBibitem
\bibitem[Goedecker(2004)]{goedecker2004minima}
S.~Goedecker, \emph{The Journal of chemical physics}, 2004, \textbf{120},
  9911--9917\relax
\mciteBstWouldAddEndPuncttrue
\mciteSetBstMidEndSepPunct{\mcitedefaultmidpunct}
{\mcitedefaultendpunct}{\mcitedefaultseppunct}\relax
\EndOfBibitem
\bibitem[Schaefer \emph{et~al.}(2015)Schaefer, Ghasemi, Roy, and
  Goedecker]{schaefer2015stabilized}
B.~Schaefer, S.~A. Ghasemi, S.~Roy and S.~Goedecker, \emph{The Journal of
  chemical physics}, 2015, \textbf{142}, 034112\relax
\mciteBstWouldAddEndPuncttrue
\mciteSetBstMidEndSepPunct{\mcitedefaultmidpunct}
{\mcitedefaultendpunct}{\mcitedefaultseppunct}\relax
\EndOfBibitem
\bibitem[Roy \emph{et~al.}(2008)Roy, Goedecker, and Hellmann]{roy2008bell}
S.~Roy, S.~Goedecker and V.~Hellmann, \emph{Physical Review E}, 2008,
  \textbf{77}, 056707\relax
\mciteBstWouldAddEndPuncttrue
\mciteSetBstMidEndSepPunct{\mcitedefaultmidpunct}
{\mcitedefaultendpunct}{\mcitedefaultseppunct}\relax
\EndOfBibitem
\bibitem[Schaefer \emph{et~al.}(2014)Schaefer, Mohr, Amsler, and
  Goedecker]{schaefer2014minima}
B.~Schaefer, S.~Mohr, M.~Amsler and S.~Goedecker, \emph{The Journal of chemical
  physics}, 2014, \textbf{140}, 214102\relax
\mciteBstWouldAddEndPuncttrue
\mciteSetBstMidEndSepPunct{\mcitedefaultmidpunct}
{\mcitedefaultendpunct}{\mcitedefaultseppunct}\relax
\EndOfBibitem
\bibitem[Sadeghi \emph{et~al.}(2013)Sadeghi, Ghasemi, Schaefer, Mohr, Lill, and
  Goedecker]{sadeghi2013metrics}
A.~Sadeghi, S.~A. Ghasemi, B.~Schaefer, S.~Mohr, M.~A. Lill and S.~Goedecker,
  \emph{The Journal of chemical physics}, 2013, \textbf{139}, 184118\relax
\mciteBstWouldAddEndPuncttrue
\mciteSetBstMidEndSepPunct{\mcitedefaultmidpunct}
{\mcitedefaultendpunct}{\mcitedefaultseppunct}\relax
\EndOfBibitem
\bibitem[Genovese \emph{et~al.}(2008)Genovese, Neelov, Goedecker, Deutsch,
  Ghasemi, Willand, Caliste, Zilberberg, Rayson,
  Bergman,\emph{et~al.}]{genovese2008daubechies}
L.~Genovese, A.~Neelov, S.~Goedecker, T.~Deutsch, S.~A. Ghasemi, A.~Willand,
  D.~Caliste, O.~Zilberberg, M.~Rayson, A.~Bergman \emph{et~al.}, \emph{The
  Journal of chemical physics}, 2008, \textbf{129}, 014109\relax
\mciteBstWouldAddEndPuncttrue
\mciteSetBstMidEndSepPunct{\mcitedefaultmidpunct}
{\mcitedefaultendpunct}{\mcitedefaultseppunct}\relax
\EndOfBibitem
\bibitem[Hartwigsen \emph{et~al.}(1998)Hartwigsen, G{\oe}decker, and
  Hutter]{hartwigsen1998relativistic}
C.~Hartwigsen, S.~G{\oe}decker and J.~Hutter, \emph{Physical Review B}, 1998,
  \textbf{58}, 3641\relax
\mciteBstWouldAddEndPuncttrue
\mciteSetBstMidEndSepPunct{\mcitedefaultmidpunct}
{\mcitedefaultendpunct}{\mcitedefaultseppunct}\relax
\EndOfBibitem
\bibitem[Goedecker \emph{et~al.}(1996)Goedecker, Teter, and
  Hutter]{goedecker1996separable}
S.~Goedecker, M.~Teter and J.~Hutter, \emph{Physical Review B}, 1996,
  \textbf{54}, 1703\relax
\mciteBstWouldAddEndPuncttrue
\mciteSetBstMidEndSepPunct{\mcitedefaultmidpunct}
{\mcitedefaultendpunct}{\mcitedefaultseppunct}\relax
\EndOfBibitem
\bibitem[Krack(2005)]{krack2005pseudopotentials}
M.~Krack, \emph{Theoretical Chemistry Accounts}, 2005, \textbf{114},
  145--152\relax
\mciteBstWouldAddEndPuncttrue
\mciteSetBstMidEndSepPunct{\mcitedefaultmidpunct}
{\mcitedefaultendpunct}{\mcitedefaultseppunct}\relax
\EndOfBibitem
\bibitem[Willand \emph{et~al.}(2013)Willand, Kvashnin, Genovese,
  V{\'a}zquez-Mayagoitia, Deb, Sadeghi, Deutsch, and
  Goedecker]{willand2013norm}
A.~Willand, Y.~O. Kvashnin, L.~Genovese, {\'A}.~V{\'a}zquez-Mayagoitia, A.~K.
  Deb, A.~Sadeghi, T.~Deutsch and S.~Goedecker, \emph{The Journal of chemical
  physics}, 2013, \textbf{138}, 104109\relax
\mciteBstWouldAddEndPuncttrue
\mciteSetBstMidEndSepPunct{\mcitedefaultmidpunct}
{\mcitedefaultendpunct}{\mcitedefaultseppunct}\relax
\EndOfBibitem
\bibitem[Perdew \emph{et~al.}(1996)Perdew, Burke, and
  Ernzerhof]{perdew1996generalized}
J.~P. Perdew, K.~Burke and M.~Ernzerhof, \emph{Physical review letters}, 1996,
  \textbf{77}, 3865\relax
\mciteBstWouldAddEndPuncttrue
\mciteSetBstMidEndSepPunct{\mcitedefaultmidpunct}
{\mcitedefaultendpunct}{\mcitedefaultseppunct}\relax
\EndOfBibitem
\bibitem[Marques \emph{et~al.}(2012)Marques, Oliveira, and
  Burnus]{marques2012libxc}
M.~A. Marques, M.~J. Oliveira and T.~Burnus, \emph{Computer physics
  communications}, 2012, \textbf{183}, 2272--2281\relax
\mciteBstWouldAddEndPuncttrue
\mciteSetBstMidEndSepPunct{\mcitedefaultmidpunct}
{\mcitedefaultendpunct}{\mcitedefaultseppunct}\relax
\EndOfBibitem
\bibitem[Silvi \emph{et~al.}(1994)Silvi,
  Savin,\emph{et~al.}]{silvi1994classification}
B.~Silvi, A.~Savin \emph{et~al.}, \emph{Nature}, 1994, \textbf{371},
  683--686\relax
\mciteBstWouldAddEndPuncttrue
\mciteSetBstMidEndSepPunct{\mcitedefaultmidpunct}
{\mcitedefaultendpunct}{\mcitedefaultseppunct}\relax
\EndOfBibitem
\bibitem[Kresse and Furthm{\"u}ller(1996)]{kresse1996efficient}
G.~Kresse and J.~Furthm{\"u}ller, \emph{Physical review B}, 1996, \textbf{54},
  11169\relax
\mciteBstWouldAddEndPuncttrue
\mciteSetBstMidEndSepPunct{\mcitedefaultmidpunct}
{\mcitedefaultendpunct}{\mcitedefaultseppunct}\relax
\EndOfBibitem
\bibitem[De \emph{et~al.}(2014)De, Schaefer, Sadeghi, Sicher, Kanhere, and
  Goedecker]{de2014glassy}
S.~De, B.~Schaefer, A.~Sadeghi, M.~Sicher, D.~G. Kanhere and S.~Goedecker,
  \emph{Phys. Rev. Lett.}, 2014, \textbf{112}, 083401\relax
\mciteBstWouldAddEndPuncttrue
\mciteSetBstMidEndSepPunct{\mcitedefaultmidpunct}
{\mcitedefaultendpunct}{\mcitedefaultseppunct}\relax
\EndOfBibitem
\bibitem[Eyring(1935)]{eyring1935activated}
H.~Eyring, \emph{The Journal of Chemical Physics}, 1935, \textbf{3},
  107--115\relax
\mciteBstWouldAddEndPuncttrue
\mciteSetBstMidEndSepPunct{\mcitedefaultmidpunct}
{\mcitedefaultendpunct}{\mcitedefaultseppunct}\relax
\EndOfBibitem
\end{mcitethebibliography}
\bibliographystyle{rsc} 

\end{document}